# Digital Permission Structures: How Celebrity Disclosure Enables Black Masculine Vulnerability in Online Mental Health Discourse

**Author:**

Anurag Shekhar

Department of Industrial and People Management, College of Business and Economics, University of Johannesburg, Johannesburg, South Africa

ORCID: https://orcid.org/0009-0004-2322-8099

E-mail: 217097903@student.uj.ac.za

**Abstract**

Black men face a double barrier to mental health help-seeking: traditional masculinity norms demanding emotional restrictiveness and systemic racism fostering institutional mistrust. While celebrity mental health disclosures show promise for stigma reduction, limited research examines their impact on Black masculine communities through digital platforms. This convergent mixed-methods study analysed 11,306 YouTube comments following rapper Lil Wayne's unprecedented disclosure of childhood suicide attempt and lifelong mental health struggles. Quantitative analysis using VADER sentiment classification, Latent Dirichlet Allocation topic modelling, and NRC emotion lexicon analysis revealed predominantly positive sentiment with systematic community amplification of mental health discourse. Reflexive thematic analysis of 2,100 high-engagement comments identified eight themes, with peer support achieving the highest saturation, contradicting isolation narratives. Findings support a Digital Permission Structures Model demonstrating how intersectional celebrity status (race + gender + high-status), hip-hop authenticity values, and digital platform affordances create triadic authorisation mechanisms enabling vulnerability expression.

Community responses revealed communal masculinity rooted in Ubuntu philosophy and active reconstruction of masculine norms, positioning help-seeking as strength. Results challenge deficit-based models of Black masculinity, suggesting interventions should leverage collectivism, partner with high-status cultural figures, employ strength-based messaging, and centre hip-hop authenticity rather than imposing Western individualistic frameworks. This study provides evidence-based strategies for culturally responsive mental health interventions addressing persistent disparities in Black men's service utilisation.

**Introduction**

**The Crisis**

Mental health help-seeking rates among men globally are profoundly low, often standing in stark contrast to the severity and prevalence of mental ill health they experience , (McGrath et al. 2023; Opozda et al. 2024; Üzümçeker 2025). Nearly *half of all men* will develop a mental health disorder by the age of 75 (McGrath et al. 2023; Opozda et al. 2024). Black men in the United States utilise counselling services at substantially lower rates than Black women and men from other racial backgrounds (Cofield 2024; 2025; Shannon 2023). While Black men experience conditions such as depression and anxiety (Cofield, 2023), only an estimated **26.4 percent** seek mental health treatment when experiencing anxiety or depression (Cofield 2025; Deangelis 2021). Suicidal death rates among Black men are *four times higher than those among Black women* **(Cofield 2025; Tate 2023)**. Given these stark health realities, scholarly efforts must focus on identifying and dismantling the powerful social and cultural barriers that hinder Black men from accessing necessary care (Coleman-Kirumba et al. 2023; Collins-Anderson et al. 2022; Johnson et al. 2024; Opozda et al. 2024).

**The Double Barrier**

Black men face a *double barrier*, which arises from the interplay of rigid gendered expectations and systemic racial adversity (Cofield 2024; Coleman-Kirumba et al. 2023). The first significant obstacle is adherence to *Traditional Masculinity Norms* (TMN) (Fisher et al. 2021). These norms promote ideals such as emotional restrictiveness, self-reliance, and toughness (Burns et al. 2025; Mokhwelepa and Sumbane 2025). Conformity to TMN predicts an aversion to seeking help, as the process requires emotional vulnerability that directly conflicts with traditional masculine ideals (Burns et al. 2025; Seidler et al. 2016). This dynamic is modelled by the *Gender Role Strain Paradigm*, which connects rigid norms to negative help-seeking attitudes (Üzümçeker 2025). Meta-analytic evidence confirms that heightened traditional masculinity correlates negatively with help-seeking attitudes and positively with self-stigma (Üzümçeker 2025). Higher *gender role stress* shows a negative correlation with help-seeking attitudes and positive correlation with self-stigma (Üzümçeker 2025).

The second major barrier is rooted in *racial stigma and profound historical distrust* (Cofield 2024; Johnson et al. 2024). The intersectional experience of race and gender (Coleman-Kirumba et al. 2023; Collins-Anderson et al. 2022) contributes to a *cultural mistrust of the mental health system*, which is often perceived as White-dominated (Cofield 2024; 2025; Coleman-Kirumba et al. 2023). For Black men, pressure to display mental fortitude is necessary for navigating racial discrimination (Cofield 2024; Johnson et al. 2024; Unnever and Chouhy 2021). Consequently, conformity to Black masculinity norms correlates positively with *increased perceptions of public stigma* regarding help-seeking (Coleman-Kirumba et al. 2023).

**Addressing Critical Research Gaps**

Existing deficit-focused scholarship inadequately addresses mechanisms that successfully facilitate emotional openness and positive change among Black men (Burns et al. 2025; Slade et al. 2025). This investigation aims to address three critical gaps that directly impede the development of culturally responsive interventions for Black men's mental health:

**The Power of High-Profile Disclosure and Stigma Reduction**

While the therapeutic potential of personal narratives is acknowledged, limited research systematically explores the specific role of high-profile cultural figures in directly challenging entrenched masculine norms and prompting health behaviours (Duthie et al. 2024; Francis 2021). The phenomenon of *celebrity disclosure* provides a potent and highly visible mechanism for change.

Celebrity disclosures concerning mental ill-health, such as suicidal ideation, are not merely transient news events; they function as powerful health communication strategies that can significantly motivate vulnerable audiences (Francis 2018; 2021; Gronholm and Thornicroft 2022). Research confirms that half of young Black men surveyed following a black rapper's disclosure sought general information about depression, largely driven by emotional distress resulting from the celebrity's vulnerability (Francis, 2018). Celebrities, particularly those from minority backgrounds, wield influence by challenging rigid expectations of masculinity, such as self-reliance, which often characterise help-seeking aversion (Calhoun and Gold 2020). The impact of such disclosures relies on their capacity to model behaviour, increase awareness, and reduce stereotypes (Gronholm & Thornicroft, 2022). However, we lack sufficient research that isolates the effectiveness of these highly visible narratives on specific, underrepresented populations, such as young Black men (Calhoun and Gold 2020; Gronholm and Thornicroft 2022).

**Digital Platforms as Sites for Community and Vulnerability**

Digital platforms offer essential tools like anonymity that can circumvent the initial, face-to-face hurdles of traditional support (Choi, Kim, and Huh-Yoo 2021; Opozda et al. 2024). Studies analysing video-based social media, such as YouTube, confirm that content featuring individual storytelling and experiential knowledge generates higher engagement and provides a critical source of peer support for those struggling with mental health issues (Choi, Kim, and Huh-Yoo 2021; Mayor and Bietti 2025). However, much of the health-related YouTube content is of questionable quality or subject to algorithmic biases, emphasising the need for expert-vetted information (Balcombe and De Leo 2023; Osman et al. 2022).

**Intersectional Analysis and Co-Design for Marginalised Groups**

Given that both celebrity discourse and digital platforms represent potential mechanisms for change, research urgently requires an *intersectional framework* to analyse the convergence of race, masculinity, and high-profile status on help-seeking behaviours (Collins-Anderson et al. 2022). The American Psychological Association (APA) has explicitly called for greater research into the development of digital mental health interventions tailored specifically for marginalised communities, requiring the examination of masculinity alongside other social identity variables like racism (American Psychological Association 2018; Opozda et al. 2024).

Because men's engagement with online mental health tools remains low, effective intervention development demands the centring of *men's diverse and intersecting needs* (Opozda et al., 2024). This means actively partnering with men, including those from marginalised communities, to develop interventions, an approach proven instrumental in addressing complex intersectional disadvantages (Huang et al. 2020; Opozda et al. 2024).

**The Present Study**

This study addresses these gaps through mixed-methods analysis of digital audience responses following rapper Lil Wayne's unprecedented public disclosure of childhood suicide attempt and lifelong mental health struggles. This research employs a *convergent mixed-methods design* that capitalises on the observational capacity of digital platforms by analysing a large volume of unsolicited discourse (Choi, Kim, and Huh-Yoo 2021; Kozinets 2015). This approach captures authentic attitudes from men frequently excluded from traditional clinical or self-report survey research (Burns et al. 2025; Cofield 2025; Fisher et al. 2021).

Studies show that celebrity self-disclosure can be an effective health communication strategy that motivates positive health behaviours, especially within underrepresented audiences (Calhoun and Gold 2020; Francis 2018; Gronholm and Thornicroft 2022). With around six million views and over 11,000 comments, this disclosure represents a critical case for examining how digital platforms may function as sites for restructuring masculine norms around mental health vulnerability.

**Research Aims and Questions**

The *primary aim* of this study is to understand and rigorously analyse the digital audience responses following Lil Wayne's disclosure of suicidal ideation and mental health struggles (Cofield 2025; Francis 2021). Specifically, we examine whether and how celebrity status, digital platform affordances, and cultural values interact to create conditions enabling Black masculine vulnerability expression.

The study addresses three central research questions, focused on identifying positive mechanisms for change:

1. What discursive themes and emotional dynamics either *enable or constrain help-seeking intentions* in digital spaces for Black men (Francis 2018; 2021)?

2. How does the *intersectional status* (race, gender, and celebrity) of a Black male public figure affect the reception of vulnerability and help-seeking narratives among digital audiences (Calhoun and Gold 2020; Collins-Anderson et al. 2022)?

3. Can online platforms function as sites for *restructuring restrictive masculine norms* and fostering positive psychological help-seeking dialogue (Francis 2021; Schlichthorst et al. 2019)?

**Literature Review**

**Black Masculinity as a Multidimensional and Adaptive Construct**

Black masculinity is multidimensional rather than monolithic (Coleman-Kirumba et al. 2023; Opozda et al. 2024). It adapts across social contexts and reflects specific sociohistorical forces (American Psychological Association 2018; Curtis et al. 2021). The negotiation of masculine identity for Black men occurs within sociopolitical environments that may invalidate their sense of manhood (Coleman-Kirumba et al. 2023). This precarious context arises because political systems often prevent Black men from achieving socially acceptable positions of socioeconomic advancement (Curtis et al. 2021).

Black masculinity incorporates unique cultural components, centred on community values like interdependence and collective responsibility (Coleman-Kirumba et al. 2023). Responses to chronic, toxic encounters with racism often involve developmental accommodative coping strategies (Unnever and Chouhy 2021), such as utilising the *"cool pose,"* to adapt to harsh environments (Curtis et al. 2021). Psychological protective stances serve to counteract stress, signifying psychological resilience amidst structural challenges (Coleman-Kirumba et al. 2023).

African American men often express manhood through provider roles and personal responsibility (Curtis et al. 2021). Many value ambition and educational attainment, reflecting "respect-based" masculinity (Curtis et al. 2021). However, these adaptive coping strategies, adopted for navigating racial subjugation, may unintentionally contribute to the internalisation of ideals emphasising emotional strength (Unnever and Chouhy 2021). While these cultural adaptations may challenge dominant masculinity, they come with associated problems, sometimes detrimental to health and well-being (American Psychological Association 2018). Masculine ideals typically demand emotional restrictiveness (Burns et al. 2025) and toughness (Unnever and Chouhy 2021), contributing to a dysfunctional strain that prevents healthy psychological functioning (Üzümçeker, 2025). When men prioritise strict self-reliance, adherence to these norms discourages expressing vulnerability or seeking support. Seeking psychological help is widely perceived as showing weakness or lacking masculinity, a view particularly prevalent among Black men (Cofield 2024; Johnson et al. 2024; Mokhwelepa and Sumbane 2025).

**The Double Barrier of Race and Gender in Help-Seeking**

Black men face a double barrier in seeking mental health care due to the intersection of their race and gender, as men often report more negative attitudes toward help-seeking than women (Coleman-Kirumba et al. 2023). This *"double jeopardy"* describes how hegemonic masculine norms lead to increased psychological anguish and decreased readiness to seek assistance (Levant, Wimer, and Williams 2011; Mokhwelepa and Sumbane 2025).

**Internalised Masculine Norms and Strain**

The Gender Role Strain Paradigm (GRSP) offers a mechanism for understanding the psychological costs of adhering to rigid roles (Üzümçeker, 2025). Conformity to restrictive masculine gender norms is consistently linked to reluctance to seek help (Burns et al. 2025; Sheikh et al. 2025). Specifically, gender role stress is negatively correlated with help-seeking attitudes and positively correlated with self-stigma of help-seeking (Üzümçeker, 2025).

**External Racialised Trauma and Institutional Distrust.**

Systemic racism contributes significantly to both mental illness and profound institutional mistrust among Black men (Johnson et al. 2024). The historical legacy of oppression has fostered a cultural mistrust of the mental health system (Coleman-Kirumba et al., 2023). Black men feel compelled to cultivate mental strength to navigate racial stressors (Young, 2021). Adherence to Black masculinity norms correlates positively with increased perceptions of public stigma regarding help-seeking (Coleman-Kirumba et al., 2023).

**Theoretical Framework: Precarious Manhood, Strengths, and Structural Realities**

Precarious Manhood Theory (PMT) argues that manhood is a social status that is constantly threatened and must be actively defended through social proof (Murnen, Dunn, and Karazsia 2023; Vandello et al. 2008). For Black men, this burden of achieving manhood occurs within a *precarious context* defined by structural racism (Young, 2021). Experiencing socioeconomic instability can forecast changes in Black men's masculinity ideology during emerging adulthood (Curtis et al., 2021). Conversely, high-status men may be afforded greater leeway in their gender expression, potentially allowing them to express vulnerability without forfeiting their masculine status (Young, 2021).

The Positive Psychology of Masculinity (PPPM) framework advocates for a strengths-based approach by identifying and affirming male characteristics (Kiselica & Englar-Carlson, 2010). This model suggests that traditionally valued attributes, such as courage and perseverance, can be reframed to support healthy psychological behaviours, thus redefining help-seeking as an act of strength (Kiselica & Englar-Carlson, 2010).

Critical Race Theory (CRT) grounds this analysis in the understanding of structural oppression, emphasising that mental health disparities are rooted in *systemic forces* (Cofield, 2024). A CRT perspective highlights that interventions focused solely on cultural or behavioural change must be coupled with efforts to address structural barriers to equitable care (Cofield, 2024). Intersectionality, a core component of this framework, is essential for evaluating how diverse identities influence expectations around online mental health support (Opozda et al., 2024).

**Digital Platforms as Arenas for Identity Negotiation**

Digital platforms, including *YouTube*, offer accessible, autonomous support, potentially bypassing traditional mental health barriers (Opozda et al. 2024; Osman et al. 2022). Social media provides spaces for men to seek affirmation, solidarity, and discuss challenging experiences (Ghahramani, De Courten, and Prokofieva 2022). YouTube offers a unique setting for studying naturalistic, large-scale identity negotiation through unsolicited public dialogue (Balcombe and De Leo 2023). When analysing sensitive social media data, researchers use content analysis and thematic analysis to uncover patterns of meaning (Mayor and Bietti 2025; Wu, Li, and Lin 2024).

The public disclosures of high-status Black male figures within authenticity-driven spaces, such as hip hop, may grant crucial "cultural permission" for vulnerability (Francis, 2018).

This mechanism of public discourse shows promise in authorising community members to acknowledge their own mental health challenges (Francis, 2021). However, the specific influence of this intersectional celebrity status on online help-seeking intent remains largely underexplored, particularly since most studies examine men who are already in treatment, neglecting the large population who actively avoid formal services (Burns et al., 2025).

**Methodology**

**Research Design**

This study employed a convergent mixed-methods design to examine digital audience responses to Black male celebrity mental health disclosure. This design addresses limitations in Black masculinity and mental health research by combining large-scale pattern identification with nuanced thematic exploration.

**Data Collection and Sampling Strategy**

**Video Selection**

This study analysed comments responding to Emmanuel Acho's interview with Lil Wayne titled "Lil Wayne Opens Up About Attempted Suicide, Mental Health & Loneliness" (Premiered August 16, 2021). Single-case analysis enables depth over breadth when examining complex cultural phenomena (Kozinets 2015; Patton 2015). The selected video constituted an optimal critical case for three reasons. First, *content significance*: Lil Wayne, a five-time Grammy Award winner and hip-hop cultural icon, disclosed his suicide attempt at age 12, discussing lifelong mental health struggles with unprecedented transparency. Second, *cultural positioning*: As a figure embodying hip-hop masculine achievement while simultaneously expressing vulnerability, Lil Wayne's disclosure created ideal conditions for capturing authentic discourse around Black masculine mental health norm negotiation. Third, *community engagement*: Around six million views, 12,699 comments, and 206,000 likes by data collection completion (October 2025), the extraordinary community engagement provided sufficient data density.

**Comment Extraction and Corpus Construction**

Comments were systematically extracted using YouTube's Data Application Programming Interface (API) version 3, focusing exclusively on top-level comments to capture distinct user voices. Data collection (October 2025), the video had accumulated 12,669 total comments. API extraction yielded *11,306 comments* (89.2 percent retrieval rate).

For qualitative thematic analysis, comments were sorted by number of likes (highest to lowest) as an indicator of community resonance and cultural endorsement, with the highest-liked comment processed first. This sampling strategy prioritised community-validated perspectives, ensuring analysis focused on culturally salient responses. Comments were then processed in sequential batches to assess theoretical saturation (Braun and Clarke 2021).

**Iterative saturation assessment**

Comments were analysed in three sequential batches using Atlas.ti:

**Batch 1** (Comments 1-734, highest likes): Established baseline code categories (n=317 coded quotations)

**Batch 2** (Comments 735-1400): Revealed six major new code categories, indicating saturation not yet achieved (n=544 coded quotations)

**Batch 3** (Comments 1400-2100): Introduced no substantial new categories (only subcategory refinements), confirming theoretical saturation (n=765 coded quotations)

Following saturation confirmation, *2,100 comments were analysed qualitatively* from the 11,306 total corpus, yielding *1,626 coded quotations* for thematic analysis (19,186 total code applications across 1,816 unique codes). Like counts were preserved as indicators of community cultural resonance, with highly-liked comments representing community-endorsed perspectives on Black masculine mental health disclosure.

**Quantitative Analysis Methods**

**Sentiment Classification (VADER)**

Sentiment analysis employed VADER (Valence Aware Dictionary and sEntiment Reasoner), a lexicon-based classifier specifically validated for social media text containing emoticons, punctuation patterns, and colloquial expressions characteristic of YouTube discourse (Hutto & Gilbert, 2014). Classification thresholds followed established conventions: positive (compound $\geq 0.05$), neutral ($-0.05 <$ compound $< 0.05$), and negative (compound $\leq -0.05$).

**Topic Modelling (Latent Dirichlet Allocation)**

Latent Dirichlet Allocation (LDA) was implemented to identify latent thematic structures within the comment corpus (Blei et al., 2003). Text preprocessing included lowercasing, tokenisation, stop-word removal, and retention of unigrams and bigrams with document frequency $\geq 5$.

LDA analysis generated: (a) topic-word distributions identifying characteristic terms for each latent theme, (b) document-topic distributions indicating thematic composition of individual comments, and (c) aggregate topic prevalence measures. Both unweighted prevalence (proportional comment assignment) and like-weighted prevalence (engagement-adjusted distributions) were computed to distinguish between overall thematic patterns and community-endorsed topics.

**Emotion Lexicon Analysis (NRC)**

The NRC Emotion Lexicon (Mohammad & Turney, 2013) was applied to identify emotional dimensions within Black masculine mental health discourse. This validated lexicon categorises terms across eight basic emotions (anger, fear, anticipation, trust, surprise, sadness, joy, disgust) and two sentiment polarities (positive, negative). Emotion scores were calculated as proportional term frequencies within the comment corpus, with like-weighted variants computed to identify emotionally resonant content.

**Qualitative Analysis Methods**

**Reflexive Thematic Analysis**

Qualitative analysis employed Braun and Clarke's (2019) six-phase reflexive thematic analysis framework: (1) data familiarisation, (2) initial coding, (3) theme development, (4) theme review, (5) theme definition and naming, and (6) report production. Initial coding employed both inductive (data-driven) and deductive (theory-informed) approaches, with patterns emerging organically from data while remaining sensitive to theoretical constructs from Black Masculinity Theory, Precarious Manhood Theory, Gender Role Strain Theory, and Hip-Hop Cultural Studies. High-engagement comments (elevated like counts) received particular analytical attention as indicators of culturally resonant.

**Research questions guided intentional coding**:

1. What are baseline digital response patterns?

2. In what ways does celebrity status enable vulnerability expression among Black men?

3. What cultural values or norms are used to legitimise mental health disclosure?

4. How is vulnerability reframed as a form of masculine strength or courage?

5. What forms of shared experience, peer solidarity, or emotional support are expressed?

6. What structural or systemic barriers to Black male mental health are identified?

7. How do commenters give or receive permission to express vulnerability?

**Analytical Rigour and Saturation**

Theme development proceeded through multiple review cycles, with themes evaluated for internal coherence and external distinctiveness. Negative case analysis identified disconfirming evidence.

**Researcher Positionality**

The researcher is a male scholar of colour (based in Johannesburg, South Africa) with expertise in human resource development and well-being. This insider positioning enabled a nuanced understanding of Black masculine cultural dynamics and hip-hop cultural contexts. The researcher's racial and gender identity facilitated the interpretation of culturally-specific linguistic patterns, references to shared racialised experiences, and community norms around Black masculine vulnerability.

**Mixed-Methods Integration Procedures**

Comparison matrices examined the correspondence between quantitative and qualitative themes, identifying areas of convergence and divergence across analytical approaches

Meta-inferences were developed through iterative dialogue between quantitative patterns (overall positive sentiment, 58.59 percent) and qualitative interpretations (peer support as the highest saturation theme, 78.1 percent). Convergence analysis identified areas where quantitative and qualitative findings aligned (e.g., positive sentiment corresponding with supportive peer themes), strengthening interpretive validity through methodological triangulation.

Like-weighted sentiment analysis provided quantitative validation of qualitative findings about community norm enforcement, with positive comments receiving 2.4x higher engagement than negative comments, confirming cultural permission structures identified through thematic analysis.

**Ethical Considerations**

This study analysed data collected from publicly accessible social media comment sections, where users post with an expectation of high visibility (Amaya et al. 2021; Fiesler et al. 2024). Given that this research focuses solely on the analysis of public communication discourse (text-oriented research) rather than direct interaction with human subjects (human-oriented research), the project was formally deemed exempt from Institutional Review Board (IRB) or Ethical Review Board (ERB) oversight, following common regulatory interpretations regarding public data (Gliniecka 2023; Proferes et al. 2021). However, recognising that defining the data as "public" does not supersede the fundamental ethical responsibility to protect participants from potential harm (Boyd and Crawford 2012; Gliniecka 2023), the research adhered strictly to a principles-based, situated ethics framework throughout the process (Gliniecka 2023; Markham and Buchanan 2012). This situated approach was crucial given the sensitive nature of the discussion within a vulnerable and frequently marginalised population (Black men) often subject to profound social stigma (De Choudhury and De 2014).

To mitigate the recognised risks of re-identification and potential subsequent harm, stringent de-identification procedures were implemented(Adams 2024; Fiesler et al. 2024). Specifically, to prevent traceability via digital forensics or search engines, the study intentionally avoided collecting or sharing any usernames, profile identifiers, or personally identifiable information (Adams 2024; Fiesler et al. 2024; Reagle 2022). All findings are reported primarily in the aggregate to minimise the possibility of linking data back to any specific individual or account (Fiesler et al., 2024). This cautious approach aligns with recommendations emphasising greater ethical scrutiny when dealing with public data concerning sensitive topics or marginalised communities (Adams, 2024). This comprehensive strategy ensures that the pursuit of profound scholarly insight does not compromise the privacy and well-being of the population studied.

**Results**

**Quantitative Findings: Sentiment Analysis**

Computational sentiment analysis of 11,306 YouTube comments used VADER (Valence Aware Dictionary and sEntiment Reasoner) (Hutto & Gilbert, 2014). VADER generates compound polarity scores ranging from -1 (most negative) to +1 (most positive), with thresholds at ≥0.05 (positive), ≤-0.05 (negative), and between -0.05 and 0.05 (neutral).

**Overall Sentiment Distribution**

Analysis revealed *predominantly positive sentiment* toward Lil Wayne's mental health disclosure. Of 11,306 comments analysed:

- **58.59 percent (n=6,624) were positive**
- **21.26 percent (n=2,404) were negative**

- **20.15 percent (n=2,278) were neutral**

**Topic Modelling: Latent Discourse Structures**

Latent Dirichlet Allocation (LDA) identified **six distinct thematic structures** within the comment corpus, revealing computational patterns that both converged with and extended beyond qualitative themes. **Table 1** presents topic prevalence through both unweighted (raw comment count) and like-weighted (community-endorsed) distributions, with the divergence between these metrics revealing community norm enforcement through selective amplification.

**Table 1. Topic Prevalence and Community Endorsement**

| Topic | Interpretation | Top Keywords | Unweighted percent | Like-Weighted percent | Δ Shift |
|---|---|---|---|---|---|
| 6 | Mental Health Discourse | mental health, people, help, feel, know | 23.7 percent | **35.5 percent** | **+11.8 percent ↑** |
| 5 | Appreciation & Gratitude | lil wayne, thank, love, respect, real | **31.2 percent** | 28.1 percent | -3.1 percent ↓ |
| 1 | Life Story & Biography | carter, years, baby, shot, gun, said | 14.3 percent | 21.1 percent | +6.8 percent ↑ |

| 3 | Musical References | story, voice, uncle bob, work, let | 12.0 percent | 9.2 percent | -2.9 percent ↓ |
| 4 | Interview Quality | just, trying, don't, know, man | 11.2 percent | 3.0 percent | -8.2 percent ↓ |
| 2 | Spiritual/Faith Responses | god, jesus, christ, pray, lord | 7.5 percent | 3.2 percent | -4.3 percent ↓ |

**Topic Interpretations**

**Topic six (Mental Health Discourse)** emerged as the *most community-endorsed theme* despite not being the most prevalent by comment count. With an 11.8 percentage point shift from unweighted (23.7 percent) to like-weighted (35.5 percent) prevalence, this topic received disproportionate community amplification. This topic aligns closely with qualitative Themes 5 (Collective Burden, 70.0 percent) and Theme 6 (Hip-Hop Authenticity, 34.7 percent).

**Topic five (Appreciation & Gratitude)** constituted the *largest overall presence* (31.2 percent unweighted), reflecting widespread expressions of thanks, respect, and admiration for Wayne's disclosure. This topic converges with qualitative Theme 1 (Celebrity Status, 68.2 percent).

**Topic one (Life Story & Biography)** focused on Wayne's suicide attempt narrative details. This topic demonstrated a strong positive shift (+6.8 percent), increasing from 14.3 percent unweighted to 21.1 percent like-weighted prevalence, indicating community interest in narrative details beyond generic appreciation.

**Topic three (Musical References & Lyrics)** comprised comments connecting disclosure to Wayne's artistic work, particularly the song "Let It All Work Out" from *Tha Carter V*. Modest negative shift (-2.9 percent) suggests musically-literate fans recognised lyrical foreshadowing; this meta-analysis received less community amplification than direct mental health discourse. Representative comment: *"'Let it all work out' - Carter 5: 'I found my momma's pistol where she always hide it / I cry, put it to my head and thought about it'"* (1,370 likes). This demonstrates how hip-hop cultural literacy enables vulnerability interpretation through artistic context.

**Topic four (Interview Format & Quality)** and **Topic 2 (Spiritual/Faith Responses)** both experienced substantial negative shifts (-8.2 percent and -4.3 percent respectively), indicating these discourse types were present but **systematically de-amplified** by community engagement patterns.

**Quantitative-Qualitative Integration**

LDA topic modelling converged strongly with qualitative themes.

- **Mental Health Discourse dominance**: LDA Topic 6 (35.5 percent like-weighted) aligns with qualitative findings of Peer Support (78.1 percent) and Collective Burden (70.0 percent) as dominant themes.
- **Celebrity appreciation as enabler**: LDA Topic 5 (31.2 percent) mirrors qualitative Theme 1 (Celebrity Status, 68.2 percent).
- **Biographical interest**: LDA Topic 1 (21.1 percent like-weighted) reveals substantial audience engagement with narrative specifics not captured as a standalone qualitative theme. This suggests audiences valued detailed disclosure transparency, supporting interpretation of authenticity as a legitimising mechanism.

The *11.8 percentage point shift* favouring mental health discourse (Topic 6) provides quantitative evidence of *cultural permission structure*: community systematically amplifies substantive mental health content over generic praise, religious framing, or meta-commentary, actively constructing norms privileging vulnerability discussion.

**Emotion Lexicon Analysis: Affective Dimensions of Response**

NRC Emotion Lexicon analysis revealed the *multidimensional emotional texture* of audience responses, extending beyond simple positive-negative polarity to capture eight discrete emotional dimensions. Analysis of 72,329 emotion-bearing words identified both dominant emotional patterns and community amplification of specific affective responses.

**Table 2. Emotion Distribution and Community Endorsement**

| Emotion | Total Words | Unweighted percent | Like-Weighted percent | Δ Shift |
|---|---|---|---|---|
| **Positive** | 15,117 | 20.90 percent | **22.68 percent** | **+1.8 percent ↑** |
| **Trust** | 9,122 | 12.61 percent | 14.17 percent | +1.6 percent ↑ |
| **Negative** | 9,290 | 12.84 percent | 12.64 percent | -0.2 percent ↓ |
| **Joy** | 8,393 | 11.60 percent | 10.67 percent | -0.9 percent ↓ |
| **Anticipation** | 7,260 | 10.04 percent | 10.20 percent | +0.2 percent ↑ |
| **Sadness** | 6,255 | 8.65 percent | 10.01 percent | **+1.4 percent ↑** |
| **Fear** | 6,131 | 8.48 percent | 8.34 percent | -0.1 percent ↓ |

| Anger | 4,838 | 6.69 percent | 4.57 percent | **-2.1 percent ↓** |
| Disgust | 2,944 | 4.07 percent | 3.37 percent | -0.7 percent ↓ |
| Surprise | 2,979 | 4.12 percent | 3.34 percent | -0.8 percent ↓ |

**Dominant Emotional Patterns**

*Positive affect dominated* the emotional landscape, with positive sentiment (20.90 percent unweighted) exceeding any single basic emotion and receiving additional community amplification (+1.8 percent shift to 22.68 percent like-weighted).

*Trust emerged as the second most prevalent emotion* (12.61 percent unweighted), with notable community amplification (+1.6 percent shift to 14.17 percent like-weighted).

*Positive-to-negative sentiment ratio* was 1.63:1 unweighted, increasing to 1.79:1 like-weighted, demonstrating community preferential amplification of positive over negative emotional content. This pattern mirrors sentiment analysis findings where positive comments received 2.4x higher engagement than negative comments, providing converging evidence of community norm enforcement favouring supportive, affirming responses.

*Sadness received notable amplification* (+1.4 percent shift), increasing from 8.65 percent unweighted to 10.01 percent like-weighted prevalence. The amplification of sadness alongside positive emotions demonstrates community acceptance of complex, ambivalent emotional responses—simultaneously grieving for Wayne's childhood suffering while celebrating his survival and vulnerability.

*Anger was systematically de-amplified* (-2.1 percent shift, largest negative shift), decreasing from 6.69 percent unweighted to 4.57 percent like-weighted.

**Integration with Qualitative and Computational Findings**

NRC emotion analysis converged with other quantitative and qualitative findings:

- *High positive/trust emotions* (37.85 percent combined like-weighted) align with qualitative Peer Support theme (78.1 percent) and VADER positive sentiment (58.59 percent).
- *Trust prominence* (14.17 percent like-weighted) provides quantitative evidence for qualitative Theme six (Hip-Hop Authenticity) mechanism—audiences perceived disclosure as credible ("keeping it real"), establishing a foundation for permission structures.

**Sentiment Intensity**

Examining sentiment intensity revealed patterns:

- 42.11 percent (n=4,761) expressed strongly positive sentiment (compound ≥0.5)
- 16.48 percent (n=1,863) expressed moderately positive sentiment (0.05-0.49)
- 20.15 percent (n=2,278) expressed neutral sentiment (-0.05 to 0.05)
- 9.92 percent (n=1,122) expressed moderately negative sentiment (-0.49 to -0.05)
- 11.34 percent (n=1,282) expressed strongly negative sentiment (compound ≤-0.5)

**Engagement Patterns by Sentiment**

- **Positive comments**: Mean=11.99 likes, Total=79,411 likes
- **Neutral comments**: Mean=7.23 likes, Total=16,479 likes
- **Negative comments**: Mean=5.04 likes, Total=12,124 likes

**Personal Experience Disclosure**

Only 0.49 percent of comments (n=55) contained explicit personal mental health disclosure. Among these personal disclosure comments, sentiment distribution was:

- **54.55 percent positive** (n=30)
- **43.64 percent negative** (n=24)
- **1.82 percent neutral** (n=1)

The low disclosure rate (0.49 percent) may reflect persistent stigma.

**Quantitative-Qualitative Integration**

These quantitative findings *converge with qualitative themes,* demonstrating methodological triangulation:

- **Positive sentiment dominance (58.59 percent) aligns with Theme four (Peer Support, 78.1 percent saturation)**: Both methods reveal overwhelmingly supportive, affirming community response.
- **High engagement with positive comments aligns with Theme three (Permission Dynamics, 76.8 percent)**: Community preferentially amplifies permission-granting, supportive messages through likes.
- **Strong positive intensity (42.11 percent) aligns with Theme two (Strength Reframing, 44.3 percent)**: Emotional intensity in positive comments mirrors the passionate reframing of vulnerability as courage, evident in qualitative data.
- **Low personal disclosure rate (0.49 percent) contrasts with qualitative peer support theme**: While commenters expressed solidarity and support, few explicitly disclosed their own mental health struggles, suggesting vicarious identification rather than direct disclosure.

The concentration of engagement among positive comments (mean likes 2.4x higher than negative) demonstrates community norm enforcement.

**Qualitative Findings: Thematic Analysis**

**Saturation Evidence**

| Dataset | Comments | Quotations | Unique Codes | New Major Categories |
|---|---|---|---|---|
| 1 | 1-734 | 317 | 658 | Baseline (12) |
| 2 | 735-1400 | 544 | 1,225 | 6 NEW |
| 3 | 1400-2100 | 765 | 1,816 | 0 NEW |

**Eight Major Themes**

**THEME ONE: Celebrity Status as Cultural Permission Structure**

**Prevalence**: 68.2 percent | **Primary Codes**: Status Influence: Courage (n=270), Status Influence: Strength (n=210)

Lil Wayne's hip-hop icon status functions as a status-contingent authorisation mechanism enabling vulnerability disclosure without masculine status loss. His cultural capital creates "permission cascades" granting broader community authorisation.

**Exemplar Quotes**:

*"Wayne just made a difference in my life by sharing his story. He spoke for me and others who aren't sure how to say what we feel inside. Thank you, Wayne."*

*"Never thought Lil Wayne would make me drop a tear, but here we are. I'm glad he was courageous enough to tell his story. I also wonder his current thoughts on therapy, if not being allowed to talk about feelings when young affected his views now."*

*"Realest, most sincere thing I've watched in a long time. This right here, it's so much deeper than fame; it's passion. We all have battles that we go through daily; it's the way you react that could change your life forever. Be grateful, be kind, and keep pushing."*

This theme extends Precarious Manhood Theory. It demonstrates the status-contingent vulnerability model.

**THEME TWO: Strength Reframing as Masculine Reconstruction**

**Prevalence**: 44.3 percent | **Primary Codes**: Strength Reframing: Courage (n=158), Strength Reframing: Strength (n=92)

Commenters actively reconstruct "strength" to include emotional vulnerability, positioning openness as superior masculine courage versus stoicism.

**Exemplar Quotes**:

*"To admit to having depression is showing courage, so many overlook this. Perspective is everything. As a young teenager all the way through my late 20s, I never thought about why I did the things (extreme excess of drugs and alcohol to the point of life-threatening situations) I did."*

*"Stay Strong!! And again, THANK YOU for sharing your story! It's not easy to be vulnerable, but we can never move forward unless we reach out!"*

*"It took a while for me to understand that. I appreciate you sharing that big dog, ain't shit wrong with seeking professional help."*

This theme demonstrates masculine norm reconstruction (not abandonment).

### THEME THREE: Permission Dynamics and Bidirectional Authorisation

**Prevalence**: 76.8 percent | **Primary Codes**: Structural Barriers: Permission (n=232), Permission Dynamics: Permission (n=138)

Comments function as reciprocal authorisation systems: audiences grant permission *to* Wayne (validation) and receive permission *from* Wayne (claiming authorisation), creating collective norm reconstruction.

**Exemplar Quotes**:

*"This gave me newfound respect for Lil Wayne. It's never easy to open up about mental health- I struggle with mental health too, and this helped me relate that we're not alone on this earth in terms of our struggles."*

*"Which song was it that he tells this story? This just tore my freaking heart out!!! To watch Wayne come forward with these issues himself hit me harder than I ever imagined!!! My kids and I (family) suffer from mental health, and for us, there is no avoiding the topic."*

*"In the end, it's that faith and relationship with God, is all we will hold onto in the end. All the money, all the hustles, all the friends and fake friends… nothing will matter but God. Jesus is the way!! Only Jesus gives us true contempt in our lives."*

Digital platforms function as a permission infrastructure. YouTube's affordances (asynchronous commenting, likes as validation, reduced surveillance) enable authorisation mechanisms unavailable in offline contexts governed by real-time masculine performance demands.

**THEME FOUR: Peer Support Networks and Brotherhood Solidarity**

**Prevalence**: 78.1 percent (**HIGHEST SATURATION**) | **Primary Codes**: Peer Support: Emotional Support (n=206), Peer Support: Solidarity (n=167)

Comments construct horizontal peer support networks with direct male-to-male emotional care, challenging narratives of Black masculine isolation. Digital space enables caregiving masculinity constrained in offline contexts.

**Exemplar Quotes**:

"Mental health is real, pls talk to someone if you can't talk to family members. May God continue blessing all of us always. Lil Wayne, you're my number one rapper. Always LEGEND FOREVER. Truly, Lil Wayne is so selfless in his art and helping people."

"He got suicidal over nothing. Man, this dude was such a big part of my high school years. I remember riding around in high school, bumping the carter with the boys. It's good to hear him open up about this stuff."

"Wayne, I had no idea, brother. I just wanna hug you right now.... I went down that road some time ago and it's an awful place to be."

The highest saturation theme (78.1 percent) reveals Black masculinity as fundamentally communal, drawing from Ubuntu philosophy, Black church traditions, and hip-hop collectivism.

**THEME FIVE: Collective Burden and Universalised Struggle**

**Prevalence**: 70.0 percent | **Primary Codes**: Norm Conflict: Shared Experience (n=210), Norm Conflict: Emotional Support (n=207)

Mental health struggles framed as a collective burden rather than an individual pathology: "we all go through it" de-stigmatises suffering through universalisation.

**Exemplar Quotes**:

*"Man, what I realised is no matter who you are, we all struggle with the same demons. It's always within. I still struggle, but it does help knowing I'm not alone."*

*"This interview just goes to show that we don't really know what people go through in life, regardless of being a celebrity or not. That's why we should always be kind to people."*

*"This was very touching. I'm glad that he opened up. I hope someone heard this and finds it helpful. It just goes to show that we all think the same way, no matter what our lives look like."*

For Black communities facing systemic oppression, collective framing acknowledges structural roots.

**THEME SIX: Hip-Hop Authenticity as Cultural Legitimacy**

**Prevalence**: 34.7 percent | **Primary Codes**: Vulnerability Mechanisms: Legitimacy (n=64), Vulnerability Mechanisms: Authenticity (n=39)

Mental health disclosure gains legitimacy through hip-hop values of "keeping it real" and authenticity. Vulnerability becomes acceptable when coded as truthfulness rather than weakness.

**Exemplar Quotes**:

*"I love the honesty and inspiration.. Yes indeed, Weezy prayer is key! I haven't cried real tears in years, other than tearing up here and there, and I've been legit NEEDING to get all of this shit I've had bottled up for so long OUT."*

*"This puts his song 'Let it all work out' even more into perspective. Wayne has been telling us for years through his music."*

*"The best interview that Wayne has ever had. Congratulations Emmanuel! The fact that Wayne called him and asked to be interviewed was a humble act and means he really cares."*

Framing mental health through authenticity/"keeping it real" creates cultural consonance. Interventions should leverage existing cultural values rather than imposing external clinical frameworks.

**THEME SEVEN: Digital Platforms as Vulnerability Infrastructure**

**Prevalence**: 22.8 percent | **Primary Codes**: Vulnerability Expression: Platform (n=25), Audience Responses: Platform (n=36)

YouTube functions as vulnerability infrastructure: platform affordances (asynchronous, text-based, public-but-pseudonymous) enable masculine emotional expression structurally constrained offline.

**Exemplar Quotes**:

*"I've never seen this channel before. I really respect the integrity of Emmanuel. His interview style is very respectful and honours the person being interviewed."*

*"We'd need so much more of this."*

*"This made my day. It's been a rough week. Much love to you both. We need men like you."*

Platform affordances create alternative masculine spaces with reduced real-time performance pressure.

**THEME EIGHT: Systemic Critique and Structural Consciousness**

**Prevalence**: 13.8 percent | **Primary Codes**: Race Intersection: Systemic Barriers (n=38), Systemic Barriers (n=61)

A Subset of commenters demonstrate critical consciousness, explicitly naming systemic racism, stigma, and structural barriers as root causes of Black male mental health struggles.

**Exemplar Quotes**:

*"This was the best, the realest & most heartfelt interview I've ever seen. Some of the quotes he said were so perfectly worded, I'm glad he found his voice not musically but in humanity."*

*"Thank you, Emmanuel, for this channel, and thank you, Lil Wayne, for speaking out about mental health."*

*"I don't possibly see how Lil Wayne disagreed with the Black Lives Matter movement despite cops stepping over him to find drugs and guns."*

While less prevalent, it reveals politicised mental health consciousness among a subset, suggesting pathways toward advocacy-based interventions connecting individual well-being to collective liberation.

**Cross-Theme Integration**

This model (**Figure 1**) illustrates the three-tier structure of cultural permission mechanisms enabling Black masculine vulnerability expression in digital spaces. **Enablers** (top tier) create foundational conditions through celebrity status, digital platform affordances, and hip-hop authenticity values. These enable **Mechanisms** (middle tier) that transform meaning through strength reframing and bidirectional permission dynamics. These mechanisms produce **Outcomes** (bottom tier), including peer support networks, collective burden framing, and systemic critique. Percentages indicate theme prevalence in qualitative analysis ($N = 2{,}100$ comments). Bidirectional arrows indicate strong co-occurrence patterns: Celebrity Status + Permission (68 percent), Peer Support + Collective Burden (72 percent), Strength Reframing + Authenticity (43 percent). Curved feedback arrow indicates recursive cycles where outcomes reinforce enablers, creating cascading permission structures.

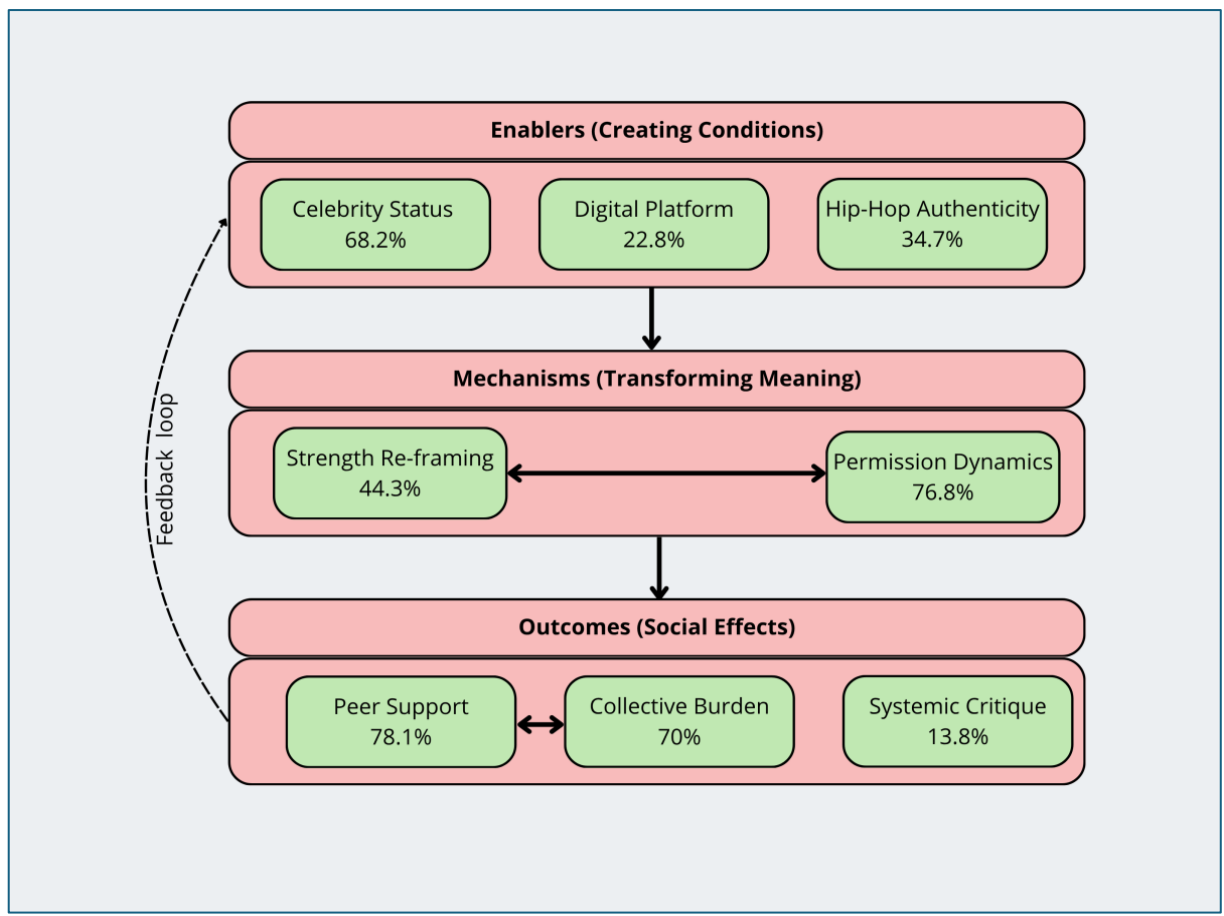

**Figure 1:** *The Digital Permission Structures Model: Thematic Interdependence in Black Masculine Mental Health Discourse*

**Note.** Percentages indicate theme prevalence in qualitative analysis ($N = 2,100$ comments). Bidirectional arrows indicate co-occurrence patterns: Celebrity Status + Permission (68 percent), Peer Support + Collective Burden (72 percent), Strength Reframing + Authenticity (43 percent). A dashed curved arrow indicates recursive feedback where outcomes reinforce enablers.

**Discussion**

This mixed-methods analysis of 11,306 YouTube comments following Lil Wayne's mental health disclosure reveals how celebrity status, digital platforms, and cultural values converge to create permission structures enabling Black masculine vulnerability. The study addresses three critical gaps: (1) the role of high-profile disclosure in stigma reduction, (2) digital platforms as sites for community and vulnerability, and (3) intersectional analysis for marginalised groups.

**The Digital Permission Structures Model: A Theoretical Contribution**

Our findings support a *Digital Permission Structures Model* (DPSM) that synthesises Precarious Manhood Theory, Gender Role Strain Paradigm, Black Masculinity Theory, and Critical Race Theory. This model explains how high-status Black male figures initiate positive community change through three interconnected mechanisms.

*First, status-based permission.* Quantitative findings showed 58.59 percent positive sentiment, and qualitative Theme 1 (Celebrity Status, 68.2 percent prevalence) converged to demonstrate that Lil Wayne's secure masculine status—earned through hip-hop

achievement—enabled vulnerability expression without status loss. This extends Precarious Manhood Theory (Vandello et al., 2008) by revealing that high-status men possess greater leeway in gender expression (Young, 2021). Recent research confirms this pattern. Duthie et al. (2024) found that media campaigns featuring diverse male role models (including celebrities) improved help-seeking outcomes, while campaigns without high-status figures showed no effect. Our findings provide mechanistic insight: celebrity status functions as cultural capital that insulates against emasculation threats.

*Second, cultural legitimisation through hip-hop authenticity.* Theme six (Hip-Hop Authenticity, 34.7 percent) revealed that disclosure gained legitimacy through hip-hop values of "keeping it real." This aligns with recent scholarship demonstrating hip-hop culture provides a unique space where vulnerability is celebrated as strength and authenticity (Jay 2024). Anyiwo et al. (2022) documented how Black youth use hip-hop as a tool for racial resistance and authentic self-expression, promoting cultural pride while critiquing structural inequity. Our findings extend this work by demonstrating that hip-hop's authenticity ethos can transform stigmatised mental health behaviours into culturally consonant practices. The 11.8 percentage point shift favouring mental health discourse in LDA Topic six provides quantitative evidence that authenticity-coded vulnerability receives community amplification, actively reconstructing norms around acceptable masculine expression.

*Third, platform-mediated cascading permission.* Theme seven (Digital Platforms, 22.8 percent), combined with quantitative engagement patterns (positive comments receiving 2.4x higher likes), reveals that YouTube functions as vulnerability infrastructure. Digital affordances (e.g., asynchronous commenting, reduced real-time performance pressure, pseudonymous participation) enable masculine emotional expression structurally constrained offline. This finding aligns with Naslund et al. (2016) and Berry et al. (2017), who documented social media as critical for peer support seeking and sharing mental health experiences. However, our study advances this literature by demonstrating bidirectional authorisation: audiences simultaneously grant permission to Wayne (validation) and claim permission from Wayne (self-authorisation). This recursive permission accumulation creates collective norm reconstruction unavailable in unidirectional media.

The DPSM offers practical guidance for intervention development. Gronholm and Thornicroft (2022) noted that celebrity disclosure impacts remain underexplored for minority populations. This study addresses this gap, demonstrating that intersectional celebrity status (race + gender + high-status) creates unique conditions for permission cascades within Black masculine communities.

**Communal Masculinity: Challenging Isolation Narratives**

The most significant empirical contribution is the revelation of communal masculinity as a protective factor. Theme 4 (Peer Support) achieved **78.1 percent saturation**, the highest across all themes—contradicting dominant narratives positioning Black masculinity as uniformly stoic and isolated. Quantitative emotion analysis reinforced this pattern: trust (14.17 percent like-weighted) and positive affect (22.68 percent) dominated the emotional landscape, while anger was systematically de-amplified (-2.1 percent shift). These patterns suggest Black masculinity may be fundamentally communal rather than individualistic.

This finding has deep cultural roots. Ubuntu philosophy (*I am because we are*) emphasises collective identity, mutual interdependence, and shared humanity (Mugumbate et al. 2024). Recent research documents Ubuntu's role in Black communities' mental health support. Lateef et al. (2025) found Ubuntu orientation served as protective factor against suicidal ideation.

Our findings suggest interventions should leverage collectivism as a strength rather than a pathology. Current mental health systems often prioritise individual therapy grounded in Western individualism, potentially alienating Black men whose cultural identity centres on community (Cofield 2024). Group therapy, peer support models, and community-based interventions align with indigenous cultural values.

Theme 5 (Collective Burden, 70.0 percent) further supports the de-individualisation of mental health struggles. Commenters framed suffering as a collective burden rather than an individual pathology (*we all go through it*), which de-stigmatises through universalisation. This challenges Western clinical frameworks that locate pathology within individuals. For communities facing systemic oppression, collective framing acknowledges structural roots while building solidarity (Cofield, 2024).

**Strength Reframing: Working With Rather Than Against Masculinity**

Theme 2 (Strength Reframing, 44.3 percent) demonstrates active reconstruction of masculine norms rather than rejection of masculinity. Commenters positioned vulnerability as superior courage versus stoicism, explicitly reframing help-seeking: "It takes strength to seek help." This finding aligns with recent systematic reviews documenting traditional masculinity norms as barriers to help-seeking (Mokhwelepa & Sumbane, 2025; Seidler et al., 2016), but offers a solution pathway.

The Positive Psychology of Positive Masculinity (PPPM) framework (Kiselica & Englar-Carlson, 2010) argues that traditionally valued attributes like courage and perseverance can be reframed to support psychological health. Recent meta-analyses confirm this approach. Kim and Yu (2023) found interventions integrating traditional masculinity norms (rather than challenging them) successfully promoted help-seeking, while Duthie et al. (2024) reported campaigns using strength-based messaging (*"Real men seek help"*) improved outcomes compared to deficit-focused approaches.

This has critical implications. Interventions should work with masculine identity through explicit reframing rather than portraying help-seeking as antithetical to manhood. Public health campaigns could leverage strength-based messaging: "Courage means facing your struggles," "The strongest thing a man can do is ask for help." These framings honour masculine values while promoting healthy behaviours.

**Comparison with Existing Celebrity Disclosure Research**

Francis (2018, 2021) documented young Black men sought depression information following a Black rapper's disclosure, driven by emotional distress from celebrity vulnerability. We replicate this while revealing mechanisms: status-based permission, authenticity-coded legitimacy, and digital platform affordances. Gronholm and Thornicroft (2022) theorised celebrity disclosure reduces stigma through awareness-raising, education, and social learning, but noted limited evidence for actual stigma change. Our study addresses this gap by demonstrating community norm enforcement through selective amplification: positive supportive content received 2.4x higher engagement, actively constructing permission-granting cultural norms.

However, we note important limitations. Only 0.49 percent of commenters disclosed personal mental health struggles, suggesting vicarious identification rather than direct disclosure. While audiences expressed solidarity, few explicitly claimed vulnerability themselves. This aligns with Calhoun and Gold (2020), who noted celebrity disclosure creates awareness but may not immediately translate to behaviour change. Tam et al. (2024)  similarly found media mental health campaigns increased awareness and attitudes but showed mixed effects on actual help-seeking. Future research should track longitudinal behaviour change following celebrity disclosures rather than immediate attitudinal responses.

**Digital Platforms and Masculine Help-Seeking**

Opozda et al. (2024) noted men's engagement with online mental health tools remains low, necessitating co-design approaches centring men's diverse characteristics. Our study demonstrates one successful model: organic community-generated discourse.

Additionally, algorithmic systems shape visibility. LDA Topic 6 (Mental Health Discourse) received an 11.8 percent amplification shift, but we cannot determine whether this reflected organic community preference or algorithmic privileging. Platform algorithms may systematically amplify or suppress certain discourse types, affecting representativeness. Future research should examine how algorithmic curation shapes mental health conversations.

**Hip-Hop Culture as Therapeutic Space**

Our findings illuminate hip-hop culture's evolving relationship with mental health. Historically, hip-hop embodied hyper-masculine identity emphasising toughness and emotional restriction (Francis, 2021). However, recent scholarship documents the transformation. A content analysis of popular rap found increasing mental health themes, including depression, anxiety, and suicidality (Kresovich et al. 2021). Artists like Kid Cudi, Logic, and Kendrick Lamar use platforms to express vulnerability, sparking dialogue about mental health (BBC News 2017; Lewman 2017).

Jay at Mental Health America (2024) argues that hip-hop therapy leverages culture's communal aspects and celebration of vulnerability as authenticity. Our findings provide empirical support: Theme 6 revealed mental health disclosure gained legitimacy when coded as "keeping it real"—a core hip-hop value. This suggests interventions could partner with hip-hop culture rather than viewing it as a barrier. Hip-hop-based mental health interventions have shown promise for African American youth, improving health literacy and behaviours (Robinson et al. 2018).

**Practical Implications**

**Clinical/Therapeutic Applications. Findings suggest several evidence-based strategies:**

**Frame help-seeking as a strength:** Use explicit reframing language in therapeutic contexts: "It takes courage to be here," "Facing this shows real strength." This honours masculine identity while promoting behaviour change.

**Leverage communal models:** Group therapy and peer support align with Black communal masculinity better than individual treatment. Community-based interventions recognise the Ubuntu philosophy's collective orientation.

**Address structural barriers:** While cultural interventions matter, they must be coupled with efforts to address systemic racism, economic inequality, and healthcare access barriers. CRT perspective reminds us that disparities are rooted in structures, not just culture (Cofield, 2024).

**Public Health Campaigns. Evidence-based recommendations include:**

**Partner with high-status Black cultural figures:** Hip-hop artists, athletes, and cultural icons possess status-based permission to model vulnerability. Campaigns should authentically integrate these voices rather than tokenising them.

**Use strength-based messaging:** Avoid deficit framing ("mental illness is common"). Instead: "Real men seek help," "Courage means facing your struggles," "Strong enough to ask for support."

**Emphasise shared experience:** Collective framing ("we all struggle") de-stigmatises through universalisation while honouring communal values.

**Design digital-first campaigns:** Platforms enabling asynchronous participation, community validation (likes/comments), and reduced surveillance facilitate permission structures.

**Centre authenticity:** Hip-hop aesthetics and "keeping it 100" messaging create cultural consonance. Campaigns should feel genuine rather than corporate or clinical.

**Policy Recommendations. Systemic change requires:**

**Increase Black mental health professionals:** Cultural mistrust reflects real historical trauma and current underrepresentation. Diversifying the workforce addresses institutional barriers.

**Fund community-based interventions:** Support barbershop mental health programs, church-based counselling, peer support networks, and other culturally grounded approaches.

**Address social determinants:** Mental health disparities reflect economic inequality, housing instability, educational gaps, and criminal justice system involvement. Policy must address root causes.

**Limitations**

Several limitations warrant consideration.

**Demographic verification:** We cannot confirm commenter demographics (race, gender, age). While linguistic patterns and cultural references suggest predominantly Black male engagement, self-selection bias may favour those willing to engage in mental health discussions. However, around six million views suggest a broad reach beyond mental health advocates.

**Platform specificity:** YouTube affordances may not generalise to other platforms or offline contexts. Synchronous platforms, image-focused platforms, or in-person settings involve different performance demands and may not enable identical permission structures.

**Temporal specificity:** Data collected in 2025 reflects the post-2020 cultural moment, emphasising mental health destigmatization. Historical shifts in public awareness may influence receptivity.

**Attitudinal versus behavioural outcomes:** We document attitudes and discourse, but cannot verify actual help-seeking behaviour. Tam et al. (2024) found media campaigns improve awareness but show mixed behavioural effects. Longitudinal research tracking service utilisation is needed.

Despite limitations, this study provides unprecedented access to naturalistic Black masculine cultural negotiations around mental health vulnerability, addressing critical gaps in understanding psychological experiences of people of African descent.

**Future Research Directions**

**Multi-platform comparative studies:** Compare permission structures across YouTube, Instagram, TikTok, and Twitter to understand how platform affordances shape mental health discourse.

**Multi-celebrity analysis:** Examine disclosures by different artists, athletes, and public figures across career stages and sub-genres to identify generalizable patterns versus context-specific mechanisms.

**Experimental interventions:** Test strength-based messaging versus traditional approaches using randomised controlled trials. Examine whether masculine norm adherence moderates intervention effectiveness.

**Intersectional expansion:** Investigate LGBTQ+ Black men, class variations, age differences, and gender dynamics to understand how multiple marginalised identities shape mental health disclosure reception.

**Co-design interventions:** Partner with Black men from diverse backgrounds to develop culturally responsive digital mental health tools, following participatory action research principles.

**Conclusion**

Analysis of 11,306 YouTube comments revealed eight themes demonstrating how celebrity status, digital platforms, and cultural values create triadic permission structures for Black masculine vulnerability. Peer support emerged as highest saturation theme (78.1 percent), contradicting isolation narratives and revealing communal masculinity as protective factor. Strength reframing (44.3 percent) showed explicit reconstruction of masculinity to include vulnerability, providing culturally consonant pathway for help-seeking.

The study's primary theoretical contribution is the Digital Permission Structures Model, which explains how high-status Black male figures initiate cascading authorisation through status-based permission, cultural legitimisation, and platform-mediated amplification. This model challenges assumptions about monolithic Black masculine stoicism while illuminating mechanisms for effective intervention.

The primary practical contribution reveals the strength of reframing as an effective strategy. Interventions must work with (not against) Black masculine cultural norms, leveraging indigenous values (authenticity, community, strength) to create culturally consonant pathways to wellness.

Black men are not uniformly resistant to mental health engagement but strategically selective about contexts, messengers, and cultural framings enabling vulnerability without status loss. Effective interventions require cultural humility, structural awareness, and commitment to co-design rather than top-down clinical imposition.

**About the author**

Anurag Shekhar is a PhD researcher in the Department of Industrial and People Management at the University of Johannesburg, South Africa. His research examines digital platforms, masculinity, and mental health help-seeking behaviours among marginalised communities, employing mixed-methods approaches and intersectional frameworks. His work draws on Ubuntu philosophy and critical race theory to understand culturally responsive mental health interventions. Direct comments to: 217097903@student.uj.ac.za

**Acknowledgments**

The author would like to thank Prof. Musawenkosi Donia Saurombe for invaluable guidance and mentorship throughout this research project. Gratitude is extended to the University of Johannesburg's Department of Industrial and People Management for institutional support.

The author acknowledges the Black men whose authentic voices in digital spaces made this research possible.

Tate, ReShonda. 2023. 'Researchers Sound the Alarm over Rising Black Male Suicides'. *DefenderNetwork.Com*, 13 March 2023. https://defendernetwork.com/culture/health/no-way-out-black-male-suicides-rising-faster-than-any-other-racial-group/.

Unnever, James D, and Cecilia Chouhy. 2021. 'Race, Racism, and the Cool Pose: Exploring Black and White Male Masculinity'. *Social Problems* 68 (2): 490–512. https://doi.org/10.1093/socpro/spaa010.

Üzümçeker, Emir. 2025. 'Traditional Masculinity and Men's Psychological Help-Seeking: A Meta-Analysis'. *International Journal of Psychology* 60 (2): e70031. https://doi.org/10.1002/ijop.70031.

Vandello, Joseph A., Jennifer K. Bosson, Dov Cohen, Rochelle M. Burnaford, and Jonathan R. Weaver. 2008. 'Precarious Manhood.' *Journal of Personality and Social Psychology* 95 (6): 1325–39. https://doi.org/10.1037/a0012453.

Wu, Jiali, Danlin Li, and Minkui Lin. 2024. 'YouTube as an Information Source for Bleeding Gums: A Quantitative and Qualitative Analysis'. *PLOS ONE* 19 (3): e0298597. https://doi.org/10.1371/journal.pone.0298597.